\documentclass{elsart}
\usepackage{amsmath}
\usepackage{graphicx}
\usepackage[english]{babel}

\begin{document}
\begin{frontmatter}
\title{MONTE CARLO SIMULATION OF VIRTUAL COMPTON SCATTERING BELOW PION THRESHOLD}
\author[gent]{P. Janssens\thanksref{emailp}},
\author[gent]{L. Van Hoorebeke\thanksref{emaill}},
\author[clermont]{H. Fonvieille},
\author[saclay]{N. D'Hose},
\author[clermont]{P.Y. Bertin},
\author[clermont]{I. Bensafa},
\author[gent]{N. Degrande},
\author[mainz]{M. Distler},
\author[roma]{R. Di Salvo},
\author[mainz]{L. Doria},
\author[mainz]{J.M. Friedrich},
\author[mainz]{J. Friedrich},
\author[virginia]{Ch. Hyde-Wright},
\author[clermont]{S. Jaminion},
\author[saclay]{S. Kerhoas},
\author[clermont]{G. Laveissi\`ere},
\author[saclay]{D. Lhuillier},
\author[orsay]{D. Marchand},
\author[mainz]{H. Merkel},
\author[saclay]{J. Roche},
\author[mainz]{G. Tamas},
\author[jefferson,william]{M. Vanderhaeghen},
\author[gent]{R. Van de Vyver},
\author[orsay]{J. Van de Wiele},
\author[mainz]{Th. Walcher}
\address[gent]{Department of Subatomic and Radiation Physics, Ghent University, 9000 Ghent, Belgium.}
\address[clermont]{LPC, Universit\'e Blaise Pascal, IN2P3, 63177 Aubiere Cedex, France.}
\address[saclay]{CEA DAPNIA-SPhN, C.E. Saclay, France.}
\address[mainz]{Institut f\"ur Kernphysik, Universit\"at Mainz, 55099 Mainz, Germany.}
\address[roma]{Istituto Nazionale di Fisica Nucleare, Sezione di Roma Tor Vergata, Italy.}
\address[virginia]{Old Dominion University, Norfolk, Virginia 23529, USA.}
\address[orsay]{Institut de Physique Nucl\'eaire d'Orsay, Universit\'e Paris-Sud 11, 91406 Orsay cedex, France.}
\address[jefferson]{Theory Center, Jefferson Lab, 12000 Jefferson Ave, Newport News, VA 23606, USA.}
\address[william]{Physics Department, The College of William \& Mary, Williamsburg, VA 23187, USA.}

\thanks[emailp]{peter@inwfsun1.ugent.be - Aspirant FWO-Flanders.}
\thanks[emaill]{luc.vanhoorebeke@ugent.be}

\begin{abstract}
This paper describes the Monte Carlo simulation developed specifically for the VCS experiments below pion threshold that have been performed at MAMI and JLab. This simulation generates events according to the (Bethe-Heitler + Born) cross-section behaviour and takes into account all relevant resolution-deteriorating effects. It determines the ``effective'' solid angle for the various experimental settings which are used for the precise determination of the photon electroproduction absolute cross section.
\end{abstract}
\end{frontmatter}

%

\section{Introduction} \label{sec:introduction}

Virtual Compton Scattering (VCS) off the nucleon $N$ is a valuable reaction to study the structure of the nucleon. VCS refers to the reaction $\gamma^*+N\rightarrow\gamma+N'$, where $\gamma^*$ and $\gamma$ represent a virtual and a real photon, respectively. It can be seen as an extension of Real Compton Scattering (RCS) to photon virtuality $Q^2 \neq 0$.  In this case six electromagnetic observables, called Generalized Polarizabilities (GPs), enter the cross section and may be determined to gain valuable insight into the structure of the scatterer. In the real-photon limit, $Q^2=0$, two of the six independent GPs are proportional to the well-known  polarizabilities $\alpha$ and $\beta$ obtained from RCS. The concept of GPs has first been worked out by Arenh\"ovel et al.~\cite{Arenhovel:1974}  for nuclei and later by Guichon et al.~\cite{Guichon:1995} for the nucleon.

VCS off the proton is studied using the $p(e,e'p')\gamma$ reaction: an electron scatters off a proton and a real photon is produced. The scattered electron and the recoiling proton are detected in coincidence, each in a high-resolution magnetic spectrometer, and real-photon production events are identified by reconstruction of the missing mass, which is zero in this reaction. The real photon can be produced either by the incoming or by the outgoing electron (the Bethe-Heitler contribution to the reaction) or by the nucleon. The nucleon contribution contains the Born part and the non-Born part. The sum of the Bethe-Heitler and the Born contributions will be denoted by BH+B. The non-Born part contains the GPs, which are accessible through the deviation of the measured $p(e,e'p')\gamma$ cross section from the BH+B cross section, the latter being perfectly calculable once the elastic form factors of the proton are known. 

The very first dedicated VCS experiment below pion threshold to obtain information on the GPs took place at the Mainz Microtron MAMI (Mainz, Germany) at $Q^2=0.33$ (GeV/c)$^2$~\cite{Roche:2000}. For the kinematics of this experiment the contribution of the GPs to the cross section had been estimated to amount to 10\%~\cite{Roche:2000}. This means that the absolute cross section had to be measured very precisely. In addition, one needed very elaborated analysis methods. The present paper is devoted to the description of the latter, which have been developed further and adapted to analyse also the next VCS experiment, performed at the Thomas Jefferson National Accelerator Laboratory JLab (Newport News, USA) at $Q^2$ values of 0.9 and 1.8 (GeV/c)$^2$~\cite{Laveissiere:2004}. Both experiments are unpolarized and they are very similar, in apparatus as well as in method. Most numerical examples given in this paper refer to the MAMI experiment.

The Monte Carlo code simulates $p(e,e'p')\gamma$ events comparable to those of the experiments. The simulation generates realistic spectra in the physical variables of interest and it has been used to determine with great accuracy what we will call effective solid angle. This effective solid angle is defined such that it does not only represent the geometrical acceptance, but it also includes the convolution of many effects. The aim of the present paper is to explain how the cross-section behaviour and the various resolution-deteriorating processes taking place in the target and in the detection systems have been taken into account. In addition the calculation of the simulated luminosity is explained. Throughout the paper the cross section used in the simulation will be often called ``VCS cross section'' for simplicity.

The paper is organized as follows: in section \ref{sec:definitions} the kinematics of the reaction, a description of the experiment and the definition of the effective solid angle are discussed.  In section \ref{sec:cross_section_and_phase_space} we outline the method used to implement the cross-section behaviour and we define the phase space in which the events are generated. Section \ref{sec:radcor} is devoted to a detailed description of the implementation of the radiative effects. Section \ref{sec:sim_pack} discusses the simulation package. Section \ref{sec:integrated_luminosity} covers the determination of the simulated luminosity and section \ref{sec:solang} the calculation of the effective solid angle. Results are presented in section \ref{sec:results}. Finally, section \ref{sec:summary} is a brief summary of the paper.



\section{Introductory definitions} \label{sec:definitions}


\subsection{The kinematics of the reaction and the experimental realization} \label{sec:kinematics}

\begin{figure}[tb]
\begin{center}
\includegraphics[width=130mm]{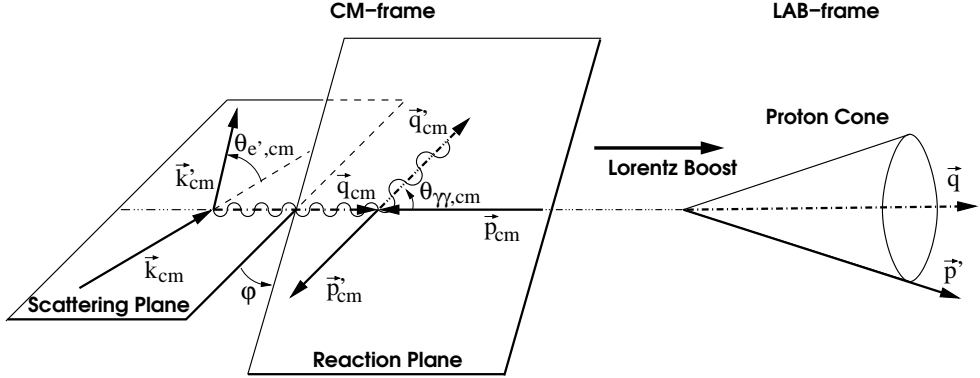}
\caption {The $p(e,e'p')\gamma$ reaction. On the left-hand side all variables are drawn in the center of mass of $\gamma^*$ and $p$. In the laboratory system the proton is boosted on a cone around $\vec{q}$ as shown on the right.}
\label{fig:kinem}
\end{center}
\end{figure}

In the process $p(e,e'p')\gamma$ an incoming electron with momentum $\vec{k}$ scatters off a proton by exchange of a virtual photon $\gamma^*$ with momentum $\vec{q}$ and a real photon with momentum $\vec{q'}$ is emitted. The vector $\vec{k}$ and the momentum vector of the outgoing electron, $\vec{k'}$, define the scattering plane. The momentum vector of the recoiling proton, $\vec{p'}$, and $\vec{q'}$ define the reaction plane. The vector $\vec{q}$, which is determined as $\vec{k}-\vec{k'}$, lies in both planes. The direction of the real photon in the CM-system of $\gamma^*$ and $p$ is determined by the angle between the two photons, $\theta_{\gamma\gamma,cm}$, and the angle $\varphi$ between the scattering and the reaction plane as is shown in figure~\ref{fig:kinem} (throughout this paper all variables in the center of mass have an index \emph{cm}; if no index is given the variable is defined in the laboratory system). $\varphi$ is defined equal to $0^{\circ}$ when $\vec{q'}$ lies in the scattering plane and points to the same side of $\vec{q}$ as $\vec{k'}$. In the CM-system, $\gamma$ and $p'$ move back to back. In the laboratory system the recoiling proton is boosted in a (narrow) cone around $\vec{q}$, while the undetected $\gamma$ can have any direction. This very welcome feature of the VCS kinematics makes it possible to cover a large range in $\theta_{\gamma\gamma,cm}$ by detecting the proton within the moderate solid angle of a high-resolution spectrometer.

In the experiment a monochromatic electron beam impinges on liquid hydrogen, contained in a metal can of known geometry. Its temperature and pressure are constantly monitored. To prevent local overheating of the liquid (which would cause density fluctuations and as such luminosity errors), the beam position on the target is continuously moving using a ``raster'' system. The scattered electron and the recoil proton are both detected in magnetic spec\-tro\-me\-ters, the entrance collimators defining their angular acceptances. The electron spectrometer defines the virtual-photon acceptance. For each electron-spectrometer setting, several proton-spectrometer settings are used to cover the interesting part of the proton cone. As both spectrometers usually rotate in the horizontal plane, one measures essentially around $\varphi=0^{\circ}$ and $\varphi=180^{\circ}$; only at sufficiently high momentum transfer and low real-photon energy the full proton cone is covered by the acceptance of the proton spectrometer.\\

%

\subsection{The solid-angle definition} 
\label{sec:basic_principles_sol}

For an ideal experiment (free of resolution effects) the cross section is determined from the number of counts detected in a given phase space bin, $N_{exp}$, and the integrated luminosity, ${\mathcal L}_{exp}$, via
\begin{eqnarray}
\frac{N_{exp}}{{\mathcal L}_{exp}} & = & \int \frac{\d \sigma}{\d\Omega}\d \Omega  = \frac{\int \frac{\d\sigma}{\d\Omega}\d\Omega}{\int\d\Omega} \int\d\Omega =  \Bigl\langle \frac{\d\sigma}{\d\Omega} \Bigr\rangle \cdot \Delta \Omega_{1},\label{eq:luminosity1}
\end{eqnarray}
where $\d\sigma/\d\Omega$ is a notation for the differential cross section and ${\d } \Omega$ represents an infinitesimal bin in the phase space under study. It is clear that in order to derive precise differential cross sections from the measured data, the solid angle of the detection apparatus has to be accurately known. Using equation~\eqref{eq:luminosity1} one determines the cross section averaged over the solid angle $\Delta\Omega_{1}$, the latter one  being a purely geometrical quantity. When the cross section has a curvature, ascribing the average cross section to the mean kinematics results in a bias. One can solve this bias by ascribing the measurement to an appropriate different kinematics (c.f.~\cite{Lafferty:1995}). This is, however, unpractical in our case because the cross section depends on five kinematical variables (see section~\ref{sec:cross_section_and_phase_space}). In this case one can stick to the central kinematics (or choose any other kinematics in the bin) and apply an appropriate correction to the average cross section in order to get an unbiased result. We choose to include this correction factor in the solid angle by defining another solid angle $\Delta \Omega_{2}$:
\begin{eqnarray}
\frac{N_{exp}}{{\mathcal L}_{exp}}& = & \Bigl( \frac{\d\sigma}{\d\Omega} \Bigr)_{0} \int \Bigl( 1 + \frac{ \frac{\d\sigma}{\d\Omega} -\bigl( \frac{\d\sigma}{\d\Omega} \bigr)_{0}}{\bigl( \frac{\d\sigma}{\d\Omega} \bigr)_{0}} \Bigr)\d\Omega \nonumber \\
 & = &  \Bigl( \frac{\d\sigma}{\d\Omega}\Bigr)_{0} \cdot (\Delta\Omega_{1}+\omega)  =  \Bigl( \frac{\d\sigma}{\d\Omega} \Bigr)_{0} \cdot \Delta \Omega_{2},\label{eq:luminosity2}
\end{eqnarray}
where $(\d \sigma/\d \Omega)_0$ is the cross section at the chosen point. The solid angle $\Delta \Omega_{2}$ deviates from $\Delta \Omega_{1}$ by the amount $\omega$, which depends on the curvature of the cross section over the bin and the chosen point in the bin. To obtain $\Delta \Omega_{2}$ one must know with sufficient accuracy the cross section behaviour of the process under study in the phase space region under consideration. In principle, this must be the cross section which one is going to measure and which is therefore unknown at the moment of the simulation. A sufficiently good approximation, however, is the BH+B cross section, since it is expected to deviate by less than 10\% from the complete $p(e,e'p')\gamma$ cross section; in particular its curvature, which is the decisive feature in this context, should be a very good approximation to the real one. 

The solid angles $\Delta\Omega_{1}$ or $\Delta\Omega_{2}$ must incorporate not only the actual detection geometry but also the various resolution effects. This is why these solid angles are called ``effective'' and why they can only be calculated by a Monte Carlo simulation.

The present simulation is used to calculate $\Delta\Omega_{2}$ of the experimental setups used in VCS experiments and, at the same time, to compare experimental and simulated data on an absolute scale. To this end, one introduces a simulated luminosity, ${\mathcal L}_{sim}$, equivalent to the experimental one. This simulated luminosity is defined by
\begin{eqnarray}
{\mathcal L}_{sim} = \frac{N'_{sim}}{\Delta \Omega' \langle \frac{\d \sigma}{\d\Omega} \rangle} \ , \label{eq:defluminositysim}
\end{eqnarray}
where $\langle \d \sigma / \d\Omega \rangle$ stands for the differential cross section in the simulation ave\-raged over a well-known solid angle, $\Delta \Omega'$, and $N'_{sim}$ is the number of events generated in $\Delta \Omega'$. Once the quantity ${\mathcal L}_{sim}$ is known, one calculates the effective solid angle $\Delta \Omega$ (which can be $\Delta \Omega_{1}$ or $\Delta \Omega_{2}$), in full parallellism with equation~\eqref{eq:luminosity1} and \eqref{eq:luminosity2} using 
\begin{eqnarray}
\Delta \Omega = \frac{N_{sim}}{{\mathcal L}_{sim} \frac{\d \sigma}{\d\Omega}} \ , \label{eq:luminositysim}
\end{eqnarray}
where $\d \sigma / \d\Omega$ is the cross section used in the simulation and $N_{sim}$ the number of events in $\Delta \Omega$. Sections \ref{sec:cross_section_and_phase_space} to \ref{sec:integrated_luminosity} describe how the various terms of this equation are obtained.


%

\section{Cross section behaviour and phase space definition} 
\label{sec:cross_section_and_phase_space}

\subsection{The implementation of the cross section behaviour}
\label{sec:cross_section_implementation}

\begin{figure}[tb]
\begin{center}
\includegraphics[width=100mm]{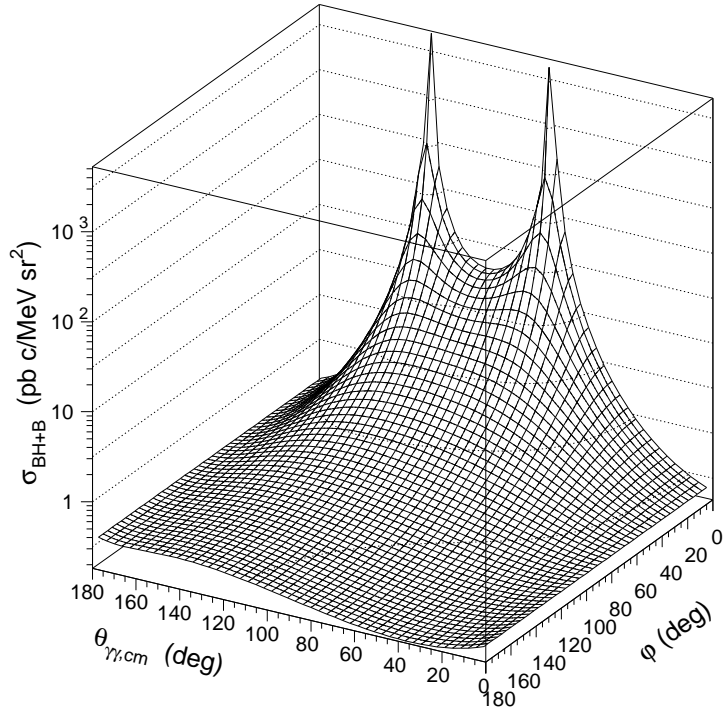}
\caption {The five-fold differential cross section for the $p(e,e'p')\gamma$ reaction as a function of $\theta_{\gamma\gamma,cm}$ and $\varphi$ ($q_{cm} = 600$ MeV/c, $q'_{cm} = 45$ MeV/c and $\varepsilon = 0.62$).}
\label{evol}
\end{center}
\end{figure}

As mentioned above, the calculation of $\Delta \Omega_2$ needs as input the cross section behaviour. The BH+B cross section, $\d ^5\sigma/\d k'\d \Omega_{e'}\d \Omega_{\gamma\gamma,cm}$, depends on the variables ($k,k',\theta_{e'},\theta_{\gamma\gamma,cm},\varphi$), where $k$, $k'$, \ldots are the moduli of the corresponding three-vectors.  Instead of $k, k'$ and $\theta_{e'}$ one can also use $q_{cm},q'_{cm}$ and the photon polarisation, $\varepsilon$, which ensures that, in the cross-section grid used by the simulation, only the real physical space is covered. An example of how the cross section behaves as a function of $\theta_{\gamma\gamma,cm}$ and $\varphi$ for fixed $q_{cm}$, $q'_{cm}$ and $\varepsilon$ for the MAMI kinematics is shown in figure~\ref{evol}. It is symmetric with respect to the scattering plane, and therefore only a ``half-sphere'' is shown. One clearly observes the two peaks corresponding to real photon emission around the incoming and outgoing electron directions. Over the complete angular range the cross section varies by orders of magnitude, but in the phase space of interest (which is away from the peak region), the cross section flattens substantially. This allows one to choose a reasonable upper limit, or envelope value, for the cross section sampling, which cuts through the peaks.
 
The number of events in an infinitesimal phase space bin is given by
\begin{equation}
\label{eq:inf_num_bins}
\d N_{bin} = {\mathcal L} \frac{\d^{5} \sigma}{\d k' \d \Omega_{e'} \d \Omega_{\gamma \gamma,cm}} 
\d k' \d\cos(\theta_{e'})\d \varphi_{e'}\d\cos(\theta_{\gamma\gamma,cm})\d\varphi ,
\end{equation}
where $\varphi_{e'}$ is the angle between the scattering plane and the horizontal plane containing the axis of the spectrometers. To generate counts in the phase space according to equation \eqref{eq:inf_num_bins} one uses the acceptance-rejection method~\cite{Yost:1988} in five dimensions with a constant as envelope for the cross section. However, the theoretical code~\cite{Vanderhaeghen:1996} used to calculate the BH+B cross section is too slow to be used on an event-by-event basis in the simulation. To solve this problem, the theoretical code has been used to calculate the BH+B cross section at the nodes of a five-dimensional grid in the variables ($q_{cm},q'_{cm},\varepsilon,\theta_{\gamma\gamma,cm},\varphi$). Then, in the simulation, the cross section value is obtained by interpolating in this grid, which makes the calculation faster by a factor of about 1000. In practice, a logarithmic interpolation is performed, reaching an accuracy of better than 1\%.\\


\subsection{The phase space definition} 
\label{sec:sim_phase_space} 

The events have to be generated according to the five-fold differential BH+B cross section in a phase-space volume $\Delta k' \cdot \Delta\Omega_{e'} \cdot  \Delta\Omega_{\gamma\gamma,cm}$. For an efficient simulation, one wants to optimize this phase space. While being not too large, it must cover the full acceptance of the apparatus, taking into account all resolution effects. The following ranges in the above mentioned variables are used:

\begin{itemize}
\item $\Delta\Omega_{e'} = \Delta cos(\theta_{e'})\cdot \Delta \varphi_{e'}$: the maximum and minimum values of $\theta_{e'}$ and $\varphi_{e'}$ are determined taking into account the shape of the extended target, the position of the spectrometer and the shape of its entrance collimator and multiple scattering effects.
\item $\Delta k'$: the lower bound is given by the lower limit of the momentum acceptance of the electron spectrometer. The upper bound is given by the maximum momentum of elastically scattered electrons in the $\Delta\Omega_{e'}$ bin defined above. This upper bound is fixed independently of the position of the elastic line relative to the electron spectrometer's momentum acceptance, since an electron, scattered with a momentum larger than the maximum accepted momentum, can still be detected due to energy losses before the spectrometer's entrance.
\item $\Delta\Omega_{\gamma\gamma,cm}$: the outgoing photon can go in any direction in the CM-system, therefore events are generated in the full solid angle $4\pi$. As a result, the outgoing proton is also sampled in its full angular phase space, i.e. the full proton cone in the laboratory. This ensures that all detectable events are indeed taken into account, even with resolution-smearing at the target. Another advantage is that the simulation can be run for several proton-spectrometer settings all at once. For each generated proton, the simulation performs a loop over the various proton-spectrometer settings and tests if the particle is accepted or not.
\end{itemize}

\begin{figure}[tb]
\begin{center}
\includegraphics[width=80mm]{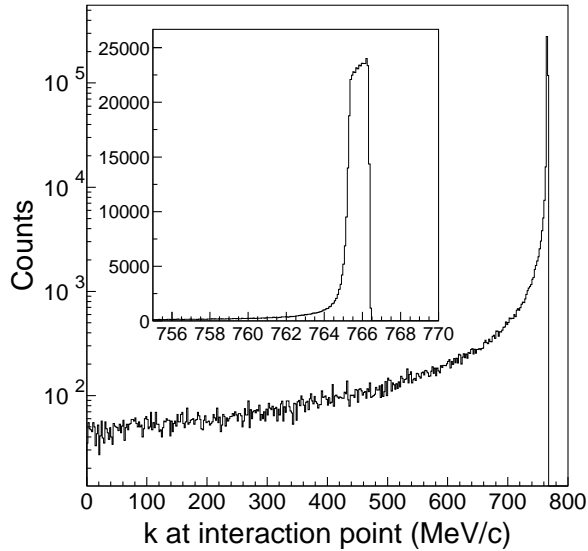}
\caption {The momentum distribution of the VCS inducing electrons as obtained by the simulation for a MAMI beam momentum of 766.4~MeV/c and $Q^2=0.33$~GeV$^2$/c$^2$. The energy losses by collision and by external and internal bremsstrahlung (before scattering) are taken into account. The insert shows a rebinned zoom on the peak region in linear scale.}
\label{fig:beam_momentum}
\end{center}
\end{figure}

The five-fold differential cross section depends on the incoming electron momentum at the interaction point, $k$, but it is not differential in this variable. However, although the beam is monochromatic (at the 10$^{-4}$ level), $k$ is not a constant. Each incoming electron loses energy in the target by collisions and by external bremsstrahlung in the material before the vertex point and by internal bremsstrahlung at the vertex point itself  (see section \ref{sec:radiative_tail}). The resulting distribution of $k$ at the interaction point is depicted on figure~\ref{fig:beam_momentum} for an incoming electron momentum of 766.4~MeV/c and a hydrogen target of 340 mg/cm$^2$.

\section{The radiation tail} \label{sec:radcor}

\subsection{The necessity to simulate a radiative tail} \label{sec:radiative_tail}

\begin{figure}[tb]
\begin{center}
\includegraphics[width=65mm]{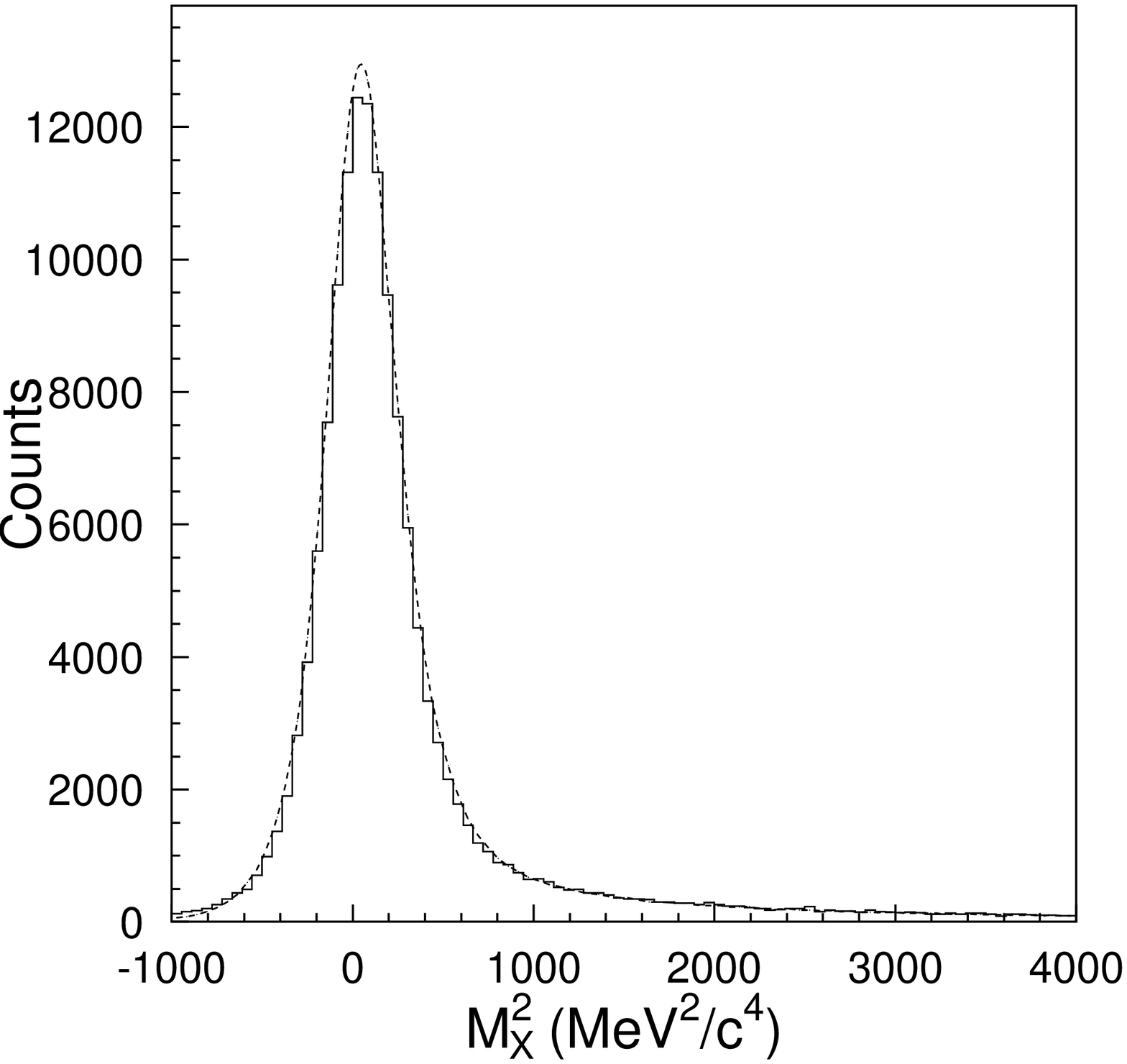}
\includegraphics[width=65mm]{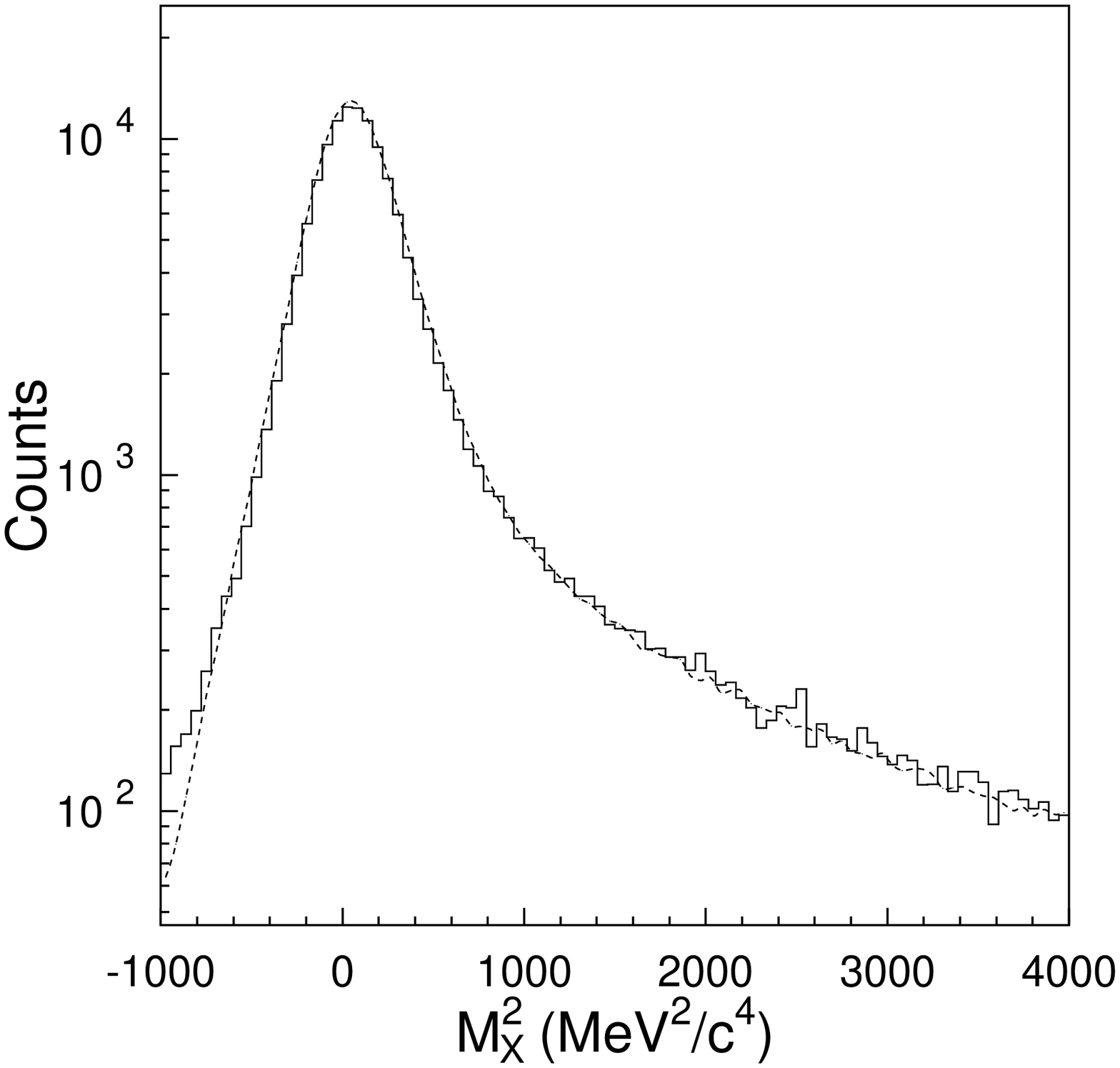}
\caption {The experimental (solid) and simulated (dashed) distributions of the missing mass squared, $ M_X^2$, for one of the MAMI kinematics ($q'_{cm} = 90$ MeV/c, $\varepsilon = 0.645$, $q_{cm} = 600$ MeV/c). For the simulation the BH+B cross section was used and the simulated distribution is normalised using the factor ${\mathcal L}_{exp}$/${\mathcal L}_{sim}$.}
\label{fig:mx2_exp_sim}
\end{center}
\end{figure}

The radiative tail is a well-known feature of electron scattering experiments: after correction for the energy losses by collisions, the energy spectrum of the scattered electron shows a peak at the kinematically expected value, but this peak is accompanied by a radiation tail to lower energies \cite{Hofstadter:1956}. This tail is due to energy loss of the incoming and outgoing electron via ionisation, external bremsstrahlung in the materials of the target and up to the spectrometer's entrances and via internal real radiation in the scattering process itself. These effects are of course also present in VCS experiments and give rise to the radiative tail observed in the spectrum of the missing mass squared $M_X^2$, defined as $({\bf k+p-k'-p'})^2$ (bold characters represent the four-vector of the particles). The resulting tail is shown in figure~\ref{fig:mx2_exp_sim}.

For the calculation of the effective solid angles, one needs a recipe to generate in the Monte Carlo simulation the radiation tail as observed. Indeed, experimentally one applies a cut in the $M_X^2$ spectrum around 0 to select real-photon production events, and the same cut must be applied to the simulated events. The simulation reproduces the radiative tail well, which is very important because one wants the final cross-section result to be independent of the cut in $M_X^2$. In fact, the influence of the position of the cut in the missing mass squared on the resulting cross section was lower than 1~\% in the MAMI case. By reproducing the radiation tail in the simulation, the part of the radiative corrections which changes the kinematics of the reaction is taken into account, and the simulated radiative tail is properly convoluted with the detector acceptance (these points will be discussed below). The other part of the radiative corrections is applied as a constant factor to the calculated cross section.

Internal and external real radiation are incorporated in the simulation. These processes are simulated by sampling in an energy-loss distribution for the incoming and outgoing electron. In the simulation only the electron's energy is changed, while its direction is assumed to be unaffected by the radiation effects (angular peaking approximation). For ionisation not only energy losses are taken into account for the electron and proton, but also multiple scattering is incorporated in the simulation. The used probability distributions are discussed in the following subsections.

\subsection{Ionisation and multiple scattering} \label{sec:ion_ms}
Collisions of the particles in the materials of the target are simulated by applying an energy loss and a scattering angle. The program {\tt glando} of the CERN-libraries \cite{CERNLIB} is used to generate a realistic energy-loss distribution based on the mean value of the energy loss, which is calculated using the Bethe-Bloch equation. The deflection caused by multiple scattering is treated as explained in \cite{Bartels:2000}.

\subsection{External radiative effects} \label{sec:ext_rad_eff}

An electron passing through a slice of material of thickness $t$ (in units of radiation length) emits photons due to brems\-strahlung. The energy loss of the electron, $\Delta E$, is equal to the sum of the energies of all produced photons. The distribution of $\Delta E$ is given in very good approximation by~\cite{Tsai:1974} ($t<0.05$) 
\begin{equation}
\label{eq:bremsstr_eloss}
I_{ext}(E_{0},\Delta E,t)=\frac{bt}{1-0.5772bt} \Bigl(\frac{\Delta E}{E_{0}}\Bigr)^{bt} \Bigl[ \frac{1}{\Delta E} \Bigl(1-\frac{\Delta E}{E_{0}}+\frac{3}{4}
(\frac{\Delta E}{E_{0}})^2 \Bigr) \Bigr].
\end{equation}
$E_{0}$ is the kinetic energy of the electron before bremsstrahlung and $b = \frac{4}{3}$.

To generate the energy loss of the electron, one samples an energy loss according to the distribution \eqref{eq:bremsstr_eloss} using the acceptance-rejection method, using $\frac{1}{\Delta E}$ as an envelope. To avoid variable overflows in the code for very small $\Delta E$, the introduction of a lower limit, $\Delta E_{ll}$, is necessary. In the present simulation $\Delta E_{ll}=1$~keV (which is well below the resolution of the experiment). Finally the electron energy is decreased by the obtained value for $\Delta E$.

In the simulation of the bremsstrahlung, only the energy of the electron is changed, which is equivalent with photon emission along the electron-momentum direction. This is a good approximation, since bremsstrahlung is very forwardly peaked. The smaller $\Delta E$, the better this approximation. Moreover, the scattering angle due to bremsstrahlung is small compared to that from multiple scattering.


\subsection{Internal radiative effects} \label{sec:internal_radiation}


\subsubsection{Virtual and Real Internal corrections}
\label{sec:int_rad_cor}

The cross section for the $p(e,e'p')\gamma$ reaction, $\sigma_{th}$, i.e. a process involving only one virtual photon and one real photon, cannot be measured directly, since in reality the pure $p(e,e'p')\gamma$ process is always accompanied by additional photons, either real or virtual. These internal radiative effects give rise to a measured cross section, $\sigma_{exp}$, which deviates from $\sigma_{th}$: 
\begin{equation}
\label{eq:int_rad_cross1}
\sigma_{exp}=(1+\delta_{tot})\sigma_{th}.
\end{equation}
The correction term $\delta_{tot}$ is negative and depends on the cut in the radiative tail accompanying the scattering process. The internal radiative corrections to VCS are discussed in great detail in~\cite{Vanderhaeghen:2000}. Written in first order, one gets 
\begin{equation}
\label{eq:int_rad_cor}
\delta^{(1)}_{tot}=\delta_{vac}+\delta_{ver}+\delta_{rad} ,
\end{equation}
$\delta_{vac}$ accounts for vacuum polarisation diagrams, $\delta_{ver}$ is the vertex correction and $\delta_{rad}$ is the correction for radiation in the one additional photon approximation. One can approximately take into account higher order radiative corrections by writing~\cite{Vanderhaeghen:2000}:
\begin{equation}
\label{eq:int_rad_cross2}
\sigma_{exp}=\frac {e^{\delta_{ver}+\delta_{rad}}}{(1-\delta_{vac}/2)^2}\sigma_{th}.
\end{equation}
For $Q^2 >> m^2$, one can write:
\begin{eqnarray}
\label{eq:deltaR}
\delta_{rad} & \approx& 
 {\frac{\alpha}{\pi}} \Bigl\{
\ln \Bigl( {\frac{ (\Delta E^c_{cm})^2}{ E_{cm} E_{cm}^{'}}} \Bigr)  
\Bigl[ \ln \Bigl(\frac {Q^2}{m^2} \Bigr) \,-\, 1 \Bigr] - {\frac{1}{2}} \ln^2 \Bigl( {\frac{E_{cm}}{E_{cm}^{'}}}  \Bigr) \, \Bigr. \nonumber\\
&&\hspace{.8cm}\Bigl. +\, \frac{1}{2} \ln^2 \Bigl( \frac{Q^2}{m^2} \Bigr) \,-\, {\frac{\pi^2}{3}} \,+\, Sp\Bigl( \cos^2 \frac{\theta_{e',cm}}{2}\Bigr) \Bigr\} \;, \\
\delta_{ver} & \approx &  \frac{\alpha}{\pi} \Bigl\{ - \frac{3}{2} \ln \Bigl(\frac {Q^2}{m^2} \Bigr)  -2 - \frac{1}{2} \ln^2\Bigl(\frac {Q^2}{m^2} \Bigr) + \frac{\pi^2}{6} \Bigr\} , \\
\delta_{vac} & \approx &  \frac{2\alpha}{3\pi} \Bigl\{ - \frac{5}{3} + \ln \Bigl(\frac {Q^2}{m^2} \Bigr)  \Bigr\} ,
\label{eq:radcorrrealbis}
\end{eqnarray}
where ${E_{cm}}$ (${E_{cm}^{'}}$) is the incoming (outgoing) electron (kinetic) energy at the reaction vertex, $\alpha$ is the fine-structure constant and $m$ is the electron mass. $Sp$ is the Spence function, eg.~\cite{Vanderhaeghen:2000}. The virtual correction terms $\delta_{ver}$ and $\delta_{vac}$ are independent of the cut in the radiative tail, $\Delta E^c_{cm}$, and nearly constant for the phase space of interest. The correction for these effects will be applied by a constant correction factor to the measured cross section. Since only the first term of $\delta_{rad}$ is dependent on $\Delta E^c_{cm}$, this term is related to the radiative tail. The other terms of  $\delta_{rad}$ are independent of the cut position and they can be considered to be constant over the phase space of interest. Therefore they will be treated in the same way as $\delta_{ver}$ and $\delta_{vac}$.

The radiative tail appears in the spectrum of the missing mass squared $M_X^2$. The cut position should be expressed in terms of $M_X^2$ since in the experiment one cuts in $M_X^2$ to identify photon-production events. The relation between  $\Delta E^c_{cm}$ and  $M_X^2$ is given by~\cite{Vanderhaeghen:2000}
\begin{equation}
\label{eq:dEs_Mx}
\Delta E^c_{cm}=\frac { \sqrt{M_X^2} }{2}.
\end{equation}
Given the relationship \eqref{eq:dEs_Mx} one could apply the correction \eqref{eq:int_rad_cross2} to obtain $\sigma_{th}$, without including the internal radiative effects in the simulation. This procedure would only be valid if the acceptance of the detectors would not cut in some parts of the phase space more severely in $M_X^2$ than the cut on the missing mass itself. This, however, is not the case in the experiments. Therefore the simulation must generate the full radiative tail by implementing electron energy losses by radiation, reproducing in this way realistic spectra.


\subsubsection{Generating a radiative tail due to internal real radiation} \label{sec:gen_rad_tail}

The first factor of the correction factor $e^{\delta_{rad}}$ is the product of a number of factors, of which the first one can be written as
\begin{equation}
\label{eq:cor_far1}
\Bigl( {\frac{ (\Delta E^c_{cm})^2}{ E_{cm} E_{cm}^{'}}} \Bigr)^{a}=
\Bigl(\frac{ \Delta E^c_{cm}}{E_{cm}} \Bigr)^a\Bigl(\frac{\Delta E^c_{cm}} {E_{cm}'} \Bigr)^a,
\end{equation}
where $a=\frac{\alpha}{\pi}\bigl[ \ln\bigl(\frac{Q^2}{m^2} \bigr)-1 \bigr]$. Assuming angular peaking, we can write~\cite{Vanderhaeghen:2000}
\begin{equation}
\label{eq:cor_far3}
\Bigl(\frac{ \Delta E^c_{cm}}{E_{cm}} \Bigr)^a\Bigl(\frac{\Delta E^c_{cm}} {E_{cm}'} \Bigr)^a= \Bigl(\frac{ \Delta E_e}{E_e} \Bigr)^a\Bigl(\frac{\Delta E_e'} { E_e'} \Bigr)^a.
\end{equation}
Following~\cite{Vanderhaeghen:2000} we interpret the factors $(\Delta E_e / E_e)^a$ and $( \Delta E_e' / E_e')^a$ as the fraction of incoming and outgoing electrons respectively, which have lost less than $\Delta E_e$ or $\Delta E'_e$ due to internal real radiation. To sample each of these energy losses $\Delta E$ one uses the distribution, $I_{int}(E,\Delta E,a)$, such that:  
\begin{equation}
\label{eq:eloss1}
\int_{0}^{\Delta E}I_{int}(E,\Delta E,a)\d(\Delta E)=\Bigl( \frac{\Delta E}{E}\Bigr)^{a} \ .
\end{equation}
Integration yields
\begin{equation}
\label{eq:eloss2}
I_{int}(E,\Delta E,a)=\frac{a}{\Delta E}\Bigl(\frac{\Delta E}{E}\Bigr)^{a},
\end{equation}
which is normalised to 1. Remark the similarity between $I_{int}(E,\Delta E,a)$ and the leading term of $I_{ext}(E,\Delta E,t)$ (eq. \eqref{eq:bremsstr_eloss}). $bt$ has been replaced by the quantity $a$, which is well known in literature as equivalent radiator \cite{Mo:1969}, i.e. an imaginary radiator placed before and after the scattering center to generate internal real radiation.

The recipe used to introduce the radiation tail due to internal radiation in the Monte Carlo simulation is:
\begin{enumerate}
\item Sample an energy loss, $\Delta E_e$, according to the distribution \eqref{eq:eloss2} with $E=$ incoming electron energy  $E_e$.
\item Generate the kinematics of a $p(e,e'p')\gamma$ event at the vertex (see figure~\ref{fig:kinem}) for the reduced energy $E_e-\Delta E_e$ of the incoming electron. The events are sampled according to the cross section at this reduced energy. After the scattering process the outgoing electron has an energy $E_e'$ at the vertex.
\item Sample an energy loss, $\Delta E_e'$, according to the distribution \eqref{eq:eloss2} with $E=E_e'$. The outgoing electron energy is now $E_e'-\Delta E_e'$.
\end{enumerate}

Remark that the above procedure implies electron-energy losses both at the incoming and the outgoing electron sides, which is fully consistent with the exponentiation idea. Again, for numerical reasons, one has to introduce a $\Delta E_{ll}$-value to sample in the $I_{int}$ distribution. In practice the sampling is done uniformly in the integrated distribution of $I_{int}$, then solving analytically for $\Delta E$. To calculate the equivalent-radiator thickness $a$, one needs the value of $Q^2$ for the event, which one can only calculate after the complete process has taken place. However, due to the slow variation of $\ln(\frac{Q^2}{m^2})$, one gets a very good approximation by using the value of $Q^2$ given by elastic electron-proton scattering at the nominal beam momentum $k_i$ and scattering angle $\theta_e$.


\section{The simulation package} \label{sec:sim_pack}

The simulation consists of three separate programs: {\tt vcssim}, {\tt resolution} and {\tt analysis}. The first one, {\tt vcssim}, generates $p(e,e'p')\gamma$ events in  the target, applying ionisation energy losses and multiple scattering to all charged particles and radiative effects to the electrons. The outgoing electron and proton are tracked up to the entrance of the spectrometers, where the collimator-acceptance cut is applied. This program produces two output files: one contains the generated events and the other one contains statistical information. The second program, {\tt resolution}, applies the resolution effects of the spectrometers on the events generated by {\tt vcssim}, producing a datafile with the events affected by the spectrometer resolution. The third program, {\tt analysis}, analyses the output datafile from {\tt resolution} in the same way experimental data are analysed and produces a third datafile. The latter contains a set of reconstructed variables to be compared to the experimental ones. The modular structure of the package has the advantage that one can change e.g. the spectrometer-resolution effects or the analysis, without having to redo the first step, which is the most time-consuming one. The three programs are described in more detail below.


\subsection{{\tt Vcssim}} \label{sec:vcssim}

Using all necessary input parameters, the program first defines the phase space in which it is going to sample (see section~\ref{sec:sim_phase_space}). In order to obtain an event the following steps are taken: first the transverse beam position on the target is generated in a horizontal and vertical distribution similar to the experimental one. An interaction point along the beamline is chosen uniformly inside the target length. The incoming electron is subject to multiple scattering, energy loss by collision and external bremsstrahlung in the target wall and the liquid hydrogen till the reaction vertex. Then the real internal radiation at the VCS vertex is simulated by an additional energy loss of the incoming electron using the equivalent-radiator approach discussed in section \ref{sec:gen_rad_tail}. Then the four-vector $\bf{k}$ of the electron inducing the actual VCS process is obtained. The energy loss through radiation can be so large, that  $k$ can already be too small to give any detectable electron in the final state. At this fixed value of $k$, the highest value of $k'$ is given by the kinematics of the elastic process $ep \to e'p'$. So at this point a test is made if the momentum of the elastically scattered electron is high enough to be accepted in the electron spectrometer. If the test is negative, the event is terminated, and a new event is generated starting all over again.

If the test is positive, one generates a scattered electron in the labframe and an outgoing real photon direction in the CM frame. The variables $\cos \theta_{e'}$,  $\varphi_{e'}$, $\cos \theta_{\gamma\gamma,cm}$,  $\varphi$ and $k'$ are all sampled uniformly in their phase space. Remark that the outgoing real photon energy is already determined by the electron kinematics. If the generated kinematics is physically possible, the cross section is calculated for this event by interpolation in the BH+B grid of section \ref{sec:cross_section_implementation}. With this value for the cross section one samples a random number between zero and the envelope value. If the value is higher than the calculated cross section, the event did not pass the acceptance-rejection test and the event is terminated.

As a next step, one has to determine whether the scattered electron and outgoing proton enter the acceptances of the spectrometers. To this end, the momenta and directions of the electron and proton have to be calculated. Based on the variables $\theta_{e'}$,  $\varphi_{e'}$ and $k'$, one can immediately calculate the four-vectors of the scattered electron and the virtual photon. Then the momentum four-vector in the center of mass for the outgoing real photon can be calculated using $\theta_{\gamma\gamma,cm}$ and  $\varphi$.  The real photon is transformed to the lab to obtain the four-vector $\bf{q'}$. The four-vector of the outgoing proton can now be calculated as $\bf{p'}=\bf{p}+\bf{k}-\bf{k'}-\bf{q'}$. The scattered electron is first subject to real internal radiation energy loss. Then it loses energy by collision and by external radiation and undergoes multiple scattering in the various materials from the vertex point to the entrance collimator of the spectrometer. Similarly, the outgoing proton will undergo energy loss by collision and multiple scattering on its way to the collimator (bremsstrahlung is negligible for such a heavy particle). Several options are proposed to calculate the collisional energy loss of particles: the mean energy loss, the most probable energy loss, or a realistic energy-loss distribution (Landau distribution). For the calculation of the effective solid angle this last option was chosen. The spectrometer acceptance is defined in different ways depending on the experiment. In the case of the MAMI experiment the angular acceptance is defined by the collimators at the entrance of the spectrometers, in the case of the JLab experiment it is defined by cuts in a five-dimensional phase space. 

The output of the {\tt vcssim} program is twofold: first a file is produced containing the events accepted by both spectrometers. One stores the kinematics at the vertex, the coordinates of the interaction point, and the momenta and angles of the particles at the spectrometer entrances. Also a proton-spectrometer index is stored, since several proton-spectrometer settings can be defined and filled simultaneously in one simulation run.  The second output file contains the simulated luminosity ${\mathcal L}_{sim}$ (see section \ref{sec:integrated_luminosity}) and some statistical information regarding the simulation run.


\subsection{{\tt Resolution}} \label{sec:resolution}

The second program, {\tt resolution}, introduces the resolution effects of the spectrometers. In the experiment, for each particle seen in the set of two double vertical drift chambers (VDCs) the trajectory, measured in the focal plane, is traced back to the target using the spectrometer optics. This yields four independent variables at the target (the momentum modulus, two projected angles and one position coordinate). The accuracy obtained on these target variables reflects the resolution of the apparatus. The program {\tt resolution} starts from the initial target variables (delivered by {\tt vcssim}) and modifies them by adding the errors. Three options to realize this are discussed below. 

As a first option one can simply make use of Gaussian-distributed resolution effects on each target variable independently, ignoring error correlations. In this scheme, the difference between the initial target variable and the modified one is sampled in a Gaussian distribution of fixed width. 

In the experiment, the resolution effects of the VDCs will cause correlations in the resolution effects on the reconstructed target variables due to the spectrometer optics. The second option reproduces these correlations. The consistency of the drift times with a straight line is used as estimate for the error on a track-by-track basis. In this way, also effects from multiple scattering within the chamber and from the used algorithm are effectively included \cite{Friedrich:2001}. After adding quadratically the contribution of multiple scattering in the spectrometer exit window, one  obtains the total error on the detector coordinates, which is propagated through the known spectrometer optics to yield the error on the target variables. From the experimental data one can fill a four-dimensional histogram for each spectrometer, where each dimension corresponds to the error on a given target variable. In this way one keeps track of error correlations (signs excluded) between the four target variables. The binning is chosen with equal width on the logarithm of the errors, which describes the distribution very precisely around the most probable value and sufficiently precisely in the long tails of the distribution, extending over four orders of magnitude, relatively to the width of the central peak. For each event the simulation samples in the four-dimensional histogram, yielding the width of the Gaussian error distribution on each target variable. Then one samples for this event in the obtained Gaussian distributions and one gets the modified target variables. This method has been applied in the analysis of the MAMI VCS experiment.

As a third option, one can implement the resolution effects in the simulation directly at the detector level. In this scheme, the accepted particle is transported to the focal plane of the spectrometer, where two types of errors are generated: 1) multiple scattering through the various materials, 2) the global resolution of the drift chambers (as deduced from experimental studies). For each particle, two tracks are considered: one with and one without these focal plane resolution effects. As in the second option, one uses the full spectrometer optics to transport the tracks back to the target. Now the quantity of interest is just the difference between the two tracks for the same particle. This difference represents the resolution effects on the target variables. Since one uses the difference between the two tracks, one can approximate the optical transport from target to focal plane: it does not need to be the exact reverse of the optical transport from focal plane to target. The method generates error correlations at the target, signs included. Large resolution tails are introduced at the level of the detector coordinates, e.g. by sampling in the sum of two Gaussian distributions with very different widths for the drift chamber resolution. This method has been applied in the analysis of the JLab VCS experiment. 

The output is a datafile containing the same variables as the one from {\tt vcssim}, but now they include also the spectrometers' resolution effects. This datafile is comparable to the experimental one. 


\subsection{{\tt Analysis}} \label{sec:analysis}

The third and final part of the simulation, the {\tt analysis} program, performs the full event reconstruction as in the analysis of the experimental data. From the reconstructed target coordinates, one first calculates the vertex point and from this the pathlengths of the particles in target materials and the corresponding (mean collisional) energy losses. The particle momenta are corrected for these energy losses, yielding the four-vectors at the vertex point. Then the complete reaction kinematics is reconstructed, including the missing particle. Then one can compare e.g. the distribution in  missing mass squared $M_X^2$ to the experimental one, as shown on figure~\ref{fig:mx2_exp_sim}.


\section{The determination of the simulated luminosity ${\mathcal L}_{sim}$} \label{sec:integrated_luminosity}

As it is clear from equation \eqref{eq:luminositysim} one needs to know the simulated luminosity in order to obtain the effective solid angle. In the experiment, the luminosity, ${\mathcal L}_{exp}$, is obtained as the product of the number of incoming electrons and the target thickness and is totally independent of the reaction under study.

The simulation uses a different approach: the luminosity in the simulation, ${\mathcal L}_{sim}$, is calculated from the cross-section samples of the acceptance-rejection method, i.e. from the reaction itself. The method is most efficient and gives a very accurate result, provided the procedure is established carefully. One counts the number of samples, $N$, generated during a simulation run in the luminosity phase space, L.P.S., which is a sub-part of the total simulation phase space (see section~\ref{sec:sim_phase_space}). Simultaneously, the cross section is integrated over this luminosity phase space. According to equation \eqref{eq:defluminositysim} ${\mathcal L}_{sim}$ is then simply given by
\begin{equation}
\label{eq:calc_luminosity}
{\mathcal L}_{sim}=\frac{N} {\int _{_{L.P.S.}} \frac{\d^{5}\sigma}{\d k' \d\Omega_{e'}\d\Omega_{\gamma\gamma,cm}}
\d k'\d\Omega_{e'}\d\Omega_{\gamma\gamma,cm}} .
\end{equation}
As such, ${\mathcal L}_{sim}$ is actually calculated in a reverse way, i.e. at the end of a simulation run, once the number of generated events is known. 

In principle one is free to define the size of the luminosity phase space. However one will have to choose a luminosity phase space that is smaller than the simulation phase space.

The first complication is due to the method used to implement the cross-section behaviour. As mentioned in section \ref{sec:cross_section_implementation}, the acceptance-rejection method with constant envelope is used, with a rejection level of about 90\%. However, among these rejected samples a large fraction can be kept to calculate the value of the cross-section integral over the luminosity phase space. One just has to make sure that the luminosity phase space does not overlap with the regions of the simulation phase space where the cross section is larger than the envelope, since the acceptance-rejection method does not work in these regions.

The second complication is connected to the fact that the real $k$ distribution has a low-momentum tail and to the fact that the cross section depends on the value of $k$. If all VCS inducing electrons would have the same momentum $k$, one could immediately apply equation \eqref{eq:calc_luminosity}. This case is explained in subsection \ref{sec:integration_phase_space}. The case of the real $k$ distribution, for which one cannot apply equation \eqref{eq:calc_luminosity} directly, is discussed in subsection \ref{sec:incoming_momentum}.


\subsection{The definition of the luminosity phase space for the case of constant $k$}
\label{sec:integration_phase_space}

\begin{figure}[tb]
\begin{center}
\includegraphics[width=85mm]{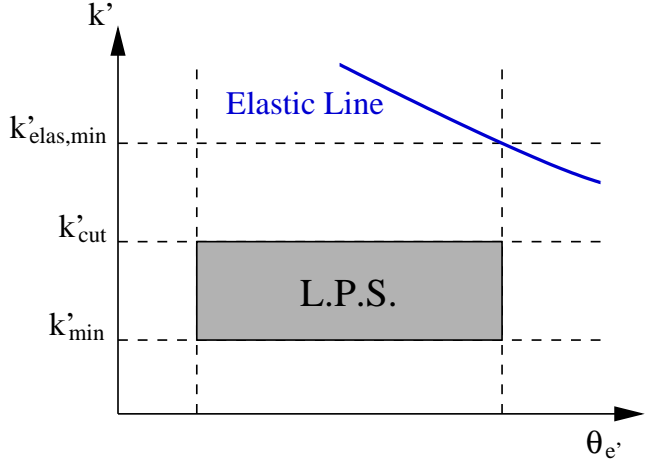}
\caption {The luminosity phase space shown schematically in the two dimensions $k'$ and $\theta_{e'}$ for the case of constant $k$. The L.P.S. is the shaded rectangle, upper-bounded by $k'_{cut}= (k'_{elas,min}+k'_{min})/2$ as explained in the text. The range in $\theta_{e'}$ is defined by the acceptance of the spectrometer. The solid line is the $(ep \to ep)$ elastic line at the incoming energy $k$. }
\label{fig:lumips}
\end{center}
\end{figure}

If all interacting electrons would have the same momentum, $k_i$ (the nominal beam momentum), the cross-section integral of equation \eqref{eq:calc_luminosity} would be given by
\begin{equation}
\label{eq:luminosity_ki}
I_{\sigma} \ = \ 
\int_{L.P.S.} \frac{\d^{5}\sigma}{\d k'
\d\Omega_{e'}\d\Omega_{\gamma\gamma,cm}}
\d k'\d\Omega_{e'}\d\Omega_{\gamma\gamma,cm}.
\end{equation}
One has to define the luminosity phase space as an integration range in $\d\Omega_{\gamma\gamma,cm}$,  $\d\Omega_{e'}$ and $\d k'$: this is done using a 5-dimensional box in $(k',$\ $\theta_{e'},$\ $\varphi_{e'},$\ $\theta_{\gamma\gamma,cm},$\ $\varphi)$, where the limits on each variable are independent of the other variables. For example, for the MAMI experiment the box has the following dimensions:
\begin{itemize}
\item In $\Delta \Omega_{\gamma\gamma,cm}$: $\theta_{\gamma\gamma,cm}$ varies from 0 to $\pi$ and $\varphi$ varies from about 0.8 to 5.48 radians. This region is chosen in order to stay away from the steep cross-section rise in the region of the BH+B peaks around $\varphi=0$ (see figure~\ref{evol}). 
\item In $\Delta \Omega_{e'}$: one uses the complete solid angle in which the electron directions are sampled.
\item In $\Delta k'$: for the minimum of $k'$, the lower limit of the electron spectrometer acceptance, $k'_{min}$, is used. For the maximum of the $k'$-integration range one has to be careful not to cross the envelope value with the cross-section values in the regions in $\Delta \Omega_{\gamma\gamma,cm}$ and  in $\Delta \Omega_{e'}$ defined above. Indeed, as  $k'$ increases at fixed $k_i$, one approaches the elastic kinematics and as such the cross section rises. To stay far enough away from the elastic kinematics, the maximum value of $k'$ for the integration range is taken equal to $k'_{cut} = (k'_{elas,min}+k'_{min})/2$. The quantity $k'_{elas,min}$ is the minimum momentum an elastically scattered electron can have in $\Delta \Omega_{e'}$, for an incoming electron momentum $k_i$. This is illustrated on figure~\ref{fig:lumips}.
\end{itemize}

\subsection{Taking into account the realistic distribution of the incoming momentum $k$} \label{sec:incoming_momentum}

Due to energy losses the electron momentum at the vertex point becomes distributed as shown in figure~\ref{fig:beam_momentum}. In this realistic case one could divide this distribution in small bins in $k$ and apply equation \eqref{eq:calc_luminosity} to calculate the partial luminosity for each bin $j$
\begin{equation}
\label{eq:lumin_bin}
{\mathcal L}_j=\frac{N_j} {I_{\sigma,j}} \ ,
\end{equation}
where $I_{\sigma,j}$ is given by expression \eqref{eq:luminosity_ki} and ${N_j}$ is the number of accepted samples, both evaluated in a luminosity phase space similar to the one of section \ref{sec:integration_phase_space}, where the range in $k$ is limited to the bin $j$. Then ${\mathcal L}_{sim}$ would be equal to $\sum_j {\mathcal L}_j$. There are two limitations to this procedure:  
\begin{enumerate}
\item It is not possible to use formula \eqref{eq:lumin_bin} on the whole incoming electron momentum distribution, because the cross-section grid does not contain cross-section samples for the entire 0 $\rightarrow$ $k_i$ incoming momentum range. One has to cut somewhere in the $k$-range, hereby defining a cut value $k_{cut}$.
\item For small values of $k$, the elastic line drawn on figure~\ref{fig:lumips} lies totally below the lower bound of the electron-momentum acceptance $k'_{min}$, therefore $k'_{cut}$ lies below $k'_{min}$ and the luminosity phase space can not be defined as in section~\ref{sec:integration_phase_space} for these electrons.
\end{enumerate}
To solve these problems the cross-section integration will be performed in a limited range of incoming electron momentum, i.e. a bin in $k$ for which the L.P.S. can be defined as in section \ref{sec:integration_phase_space}. This will yield a partial luminosity. The total luminosity will then be obtained by a simple renormalization procedure.


\subsubsection{The $k$-range for the cross section integration} \label{sec:lumi:bin_in_klab}

Quite obviously, the cross section integration of equation \eqref{eq:luminosity_ki} should be performed for the incoming electron momenta that are closest to the beam momentum $k_i$. Therefore one defines a range of the type [$k_{cut},k_i$]. The luminosity phase space is then defined as in section \ref{sec:integration_phase_space}. The value of $k'_{cut}$ is calculated using the elastic line at the lowest incoming momentum of the bin, i.e. at $k=k_{cut}$.

Of course, when one lowers the value of $k_{cut}$, one reduces the size of the luminosity phase space (due to the choice of $k'_{cut}$), and the statistical error on the luminosity increases. So one should keep  $k_{cut}$ close enough to $k_i$. For example for the MAMI experiment the range [$k_i-3$ MeV/c, $k_i$] was chosen. It contains  about 80\% of all incoming electrons, and yields a statistical error on ${\mathcal L}_{sim}$ well below 1 \%.

During execution the number of samples in the luminosity phase space, $N_{L.P.S.}$, is counted and the integral over the cross section in the L.P.S., $I_{\sigma}$, is calculated. $N_{L.P.S.}$ is the number of samples accepted by the acceptance-rejection method of section~\ref{sec:cross_section_implementation}.
At the end of execution, the partial luminosity ${\mathcal L}_{sim,3MeV/c}$ is given by $N_{L.P.S.}/I_{\sigma}$.


\subsubsection{The renormalization factor} \label{sec:luminosity_normalisation}

By the method described above we know the luminosity ${\mathcal L}_{sim,3MeV/c}$ corresponding to a fraction, $f$, of all incoming electrons, which have a momentum higher than $k_i-3$ MeV/c. This fraction $f$ is easily calculated in the simulation: one counts the total number of $k$-values that have been generated, $N_{tot}$, and the number of values that have been generated above the threshold of $k_i-3$ MeV/c, $N_{3MeV/c}$. At the end one has $f=N_{3 MeV/c}/N_{tot}$.  
However, one needs to know the total luminosity ${\mathcal L}_{sim}$ according to all incoming electrons. One can obtain the right value by correcting for the electrons one did not count in the calculation of ${\mathcal L}_{sim,3MeV/c}$. Since the luminosity is independent of the reaction under study, the total luminosity ${\mathcal L}_{sim}$ is obtained by dividing ${\mathcal L}_{sim,3MeV/c}$ by $f$.


\section{Calculation of the effective solid angle} \label{sec:solang}

The data in the output file from the {\tt analysis} program, in combination with simulated luminosity ${\mathcal L}_{sim}$ from the {\tt vcssim} program are used to calculate the effective solid angle for any given bin in the phase space, applying equation~\eqref{eq:luminositysim}:
\begin{eqnarray}
\label{eq:solid_angles}
\Delta\Omega & = & \frac{N_{sim}}
{{\mathcal L}_{sim} . \frac{\d^{5}\sigma_{sim}} {\d k'\d\Omega_{e'}\d\Omega_{\gamma\gamma,cm}}} \ ,
\end{eqnarray}
where $\d ^5\sigma_{sim}/\d k'\d \Omega_{e'}\d \Omega_{\gamma\gamma,cm}$ is now the differential cross section for the $p(e,e'p')\gamma$ reaction, used in the simulation and $N_{sim}$ the number of counts in the bin. $\Delta\Omega$ is similar as in equation \eqref{eq:luminosity1} or \eqref{eq:luminosity2}, with now the specific dimension of (sr$^2 \cdot$ MeV/c), as can be deduced from equation~\ref{eq:solid_angles}. By applying energy losses for radiative effects in the simulation, a part of the radiative correction is automatically taken into account in $\Delta\Omega$. If the cross section is taken to be a constant value over the complete phase space, equation \eqref{eq:solid_angles} will yield $\Delta\Omega_{1}$. Calculating the cross section from the data using this $\Delta\Omega_{1}$ yields the experimental cross section averaged over the bin. On the other hand, if the simulation has been performed using the BH+B cross section, the quantity $\d ^5\sigma_{sim}/\d k'\d \Omega_{e'}\d \Omega_{\gamma\gamma,cm}$ in equation \eqref{eq:solid_angles} equals the BH+B cross-section value at a given phase-space point, which can be chosen anywhere, preferentially in the bin. This procedure will give rise to  $\Delta\Omega_{2}$, comparable to equation \eqref{eq:luminosity2}. Applying this $\Delta\Omega_{2}$ to the measured data will result in the actual cross-section value at a given phase-space point. In order to get a precise result, the cross section in the simulation must have a behaviour very close to the true cross-section shape.

If necessary, one can even use an iteration procedure to improve the value of $\Delta \Omega_{2}$ by implementing in the simulation a cross section of the type BH+B plus a polarizability effect. This procedure was tested for the two experiments. In the case of MAMI, the relative change of $\Delta \Omega_2$ was smaller than 1 \%, hence the iterations had a negligible effect on the physics observables (the GPs). In the case of JLab, the relative change of $\Delta \Omega_2$ was larger, typically a few percent, translating into significant changes of the physics observables: after the first (resp. second) iteration, the GPs reached $\sim$  70 \% (resp. 90 \%) of their convergence value. The full convergence was obtained after the third iteration.


\section{Results of the effective solid angle calculation} \label{sec:results}

\begin{figure}[tb]
\begin{center}
\includegraphics[width=65mm]{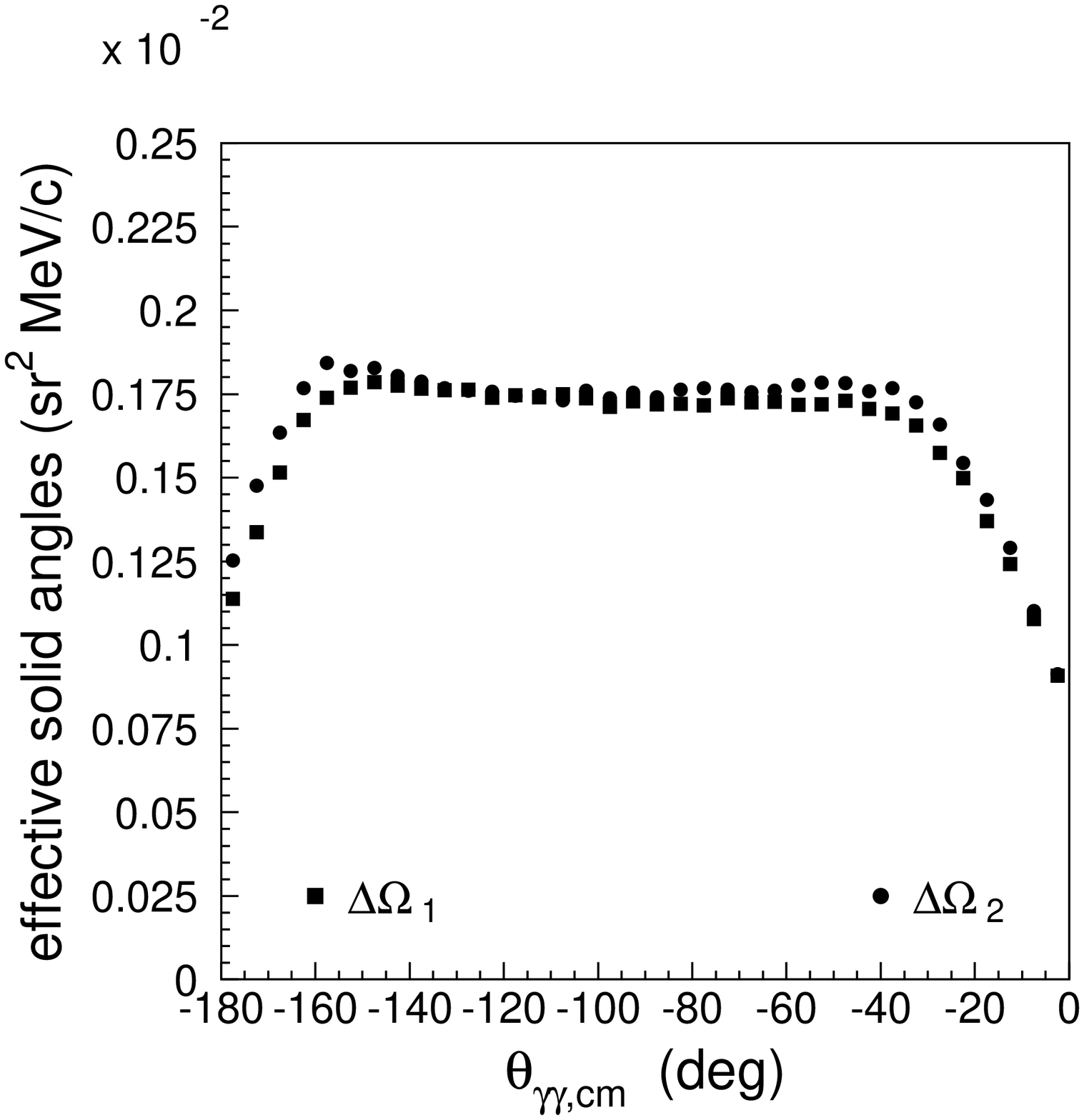}
\includegraphics[width=65mm]{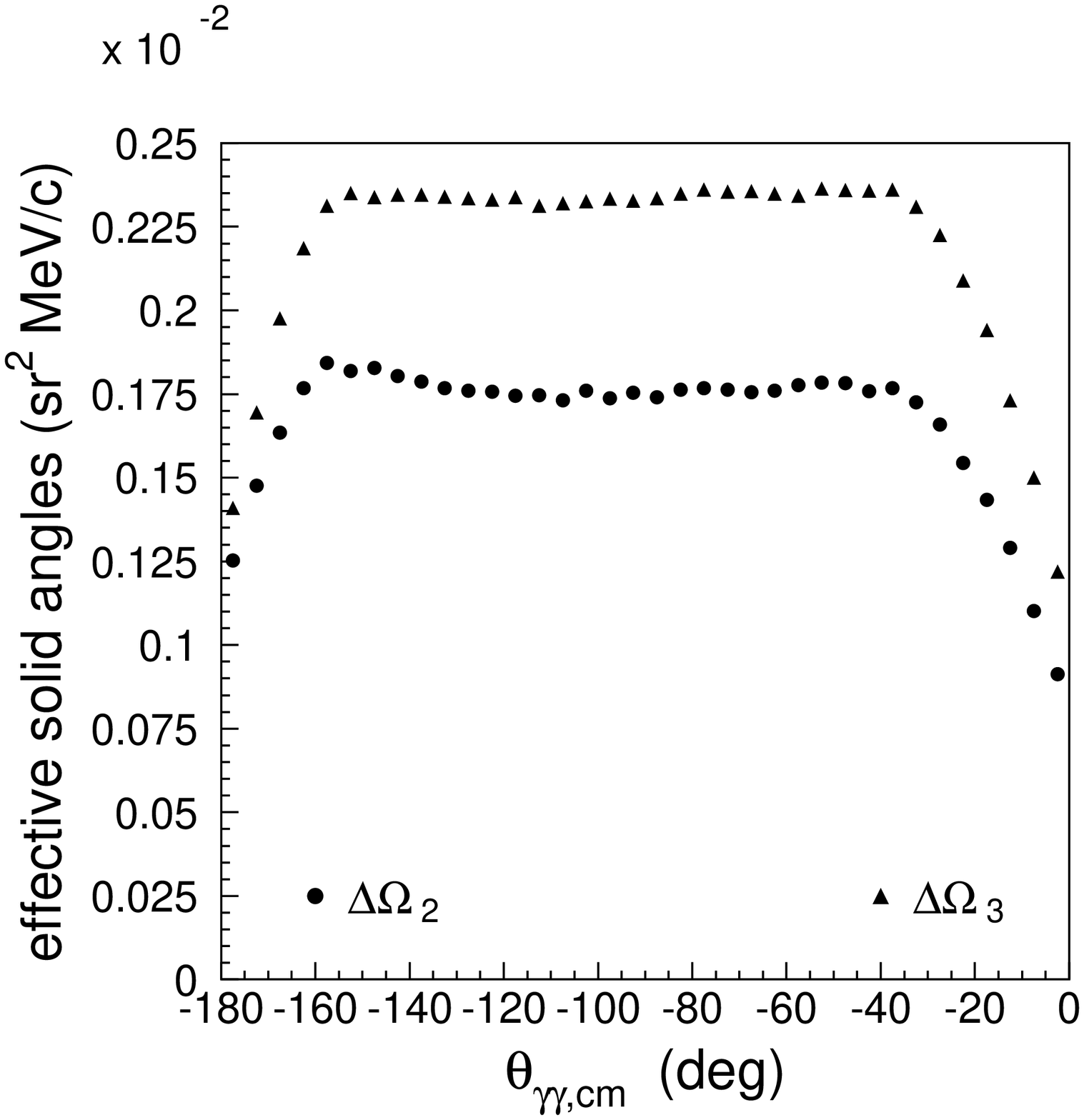}
\caption {Effective solid angles for one of the settings of the MAMI experiment as a function of $\theta_{\gamma\gamma,cm}$ ($\varphi = \pi$, $q_{cm} = 600$ MeV/c, $q'_{cm} = 45$ MeV/c and $\varepsilon = 0.62$). $\Delta\Omega_{1}$ is obtained by running the simulation with a flat cross section, for $\Delta\Omega_{2}$ the BH+B cross section is used. $\Delta\Omega_{3}$ is the same quantity as $\Delta\Omega_{2}$, but for a simulation without radiative effects. The purely statistical errors are smaller than the size of the symbols.}
\label{fig:omega12}
\end{center}
\end{figure}

As an example, figure~\ref{fig:omega12} shows the obtained effective solid angles at $q'_{cm}$ = 45 MeV/c ($\varepsilon = 0.62$ and $q_{cm} = 600$ MeV/c). The phase space is defined by 40~MeV/c $ < q'_{cm}<50$ MeV/c, $158^{\circ} <\varphi< 202^{\circ}$. The statistical error on the effective solid angle in the plateau region is about 1\%. $\Delta\Omega_{2}$ is the solid angle obtained by generating the events according to the BH+B cross section. A simulation with a constant cross section gives $\Delta\Omega_{1}$. It turns out that for this setting the difference between $\Delta\Omega_{1}$ and $\Delta\Omega_{2}$ is up to the order of 10\%. It is also interesting to run the simulation without radiative effects, which results in $\Delta\Omega_{3}$. The right panel of figure~\ref{fig:omega12} shows clearly that radiative effects have to be included in the simulation, since there is no common scaling factor between $\Delta\Omega_{2}$ and $\Delta\Omega_{3}$ for the complete phase space.

\section{Summary} \label{sec:summary}

The Monte Carlo simulation described in this paper has been developed for the analysis of the VCS experiment at MAMI and has been adapted afterwards for the analysis of the VCS experiment at JLab. It has been used to generate realistic observable spectra, which can be compared with the measured ones, and to determine accurately effective solid angles which also account for the radiative processes accompanying the VCS reaction. The use of a five-dimensional cross-section grid covering the complete simulation phase space allows to generate events according to the Bethe-Heitler+Born cross section at a very acceptable rate, using the acceptance-rejection method with a constant envelope. External and internal radiation of real photons are implemented in a well-founded way by generating realistic radiative tails and convoluting these effects with the acceptance of the detection system.

The simulation described above is flexible. All resolution deteriorating effects can independently be switched on or off and it is possible to use a constant cross section or the BH+B cross section to generate events. Due to the multiple proton-spectrometer option the yield in several proton-arm settings for one electron-spectrometer setting can be simulated in one run, while the modularity of the code gives the possibility to study spectrometer-resolution effects in an efficient way. Finally, the program is general enough to allow adaptation to many other processes, including e.g. elastic scattering and pion electroproduction.
\begin{ack}
This work was supported in part by the FWO-Flanders (Belgium), the BOF-Ghent University, the French CEA and CNRS/IN2P3, the Deutsche Forschungsgemeinschaft (SFB 201 and SFB 443) and by the Federal State of Rhineland-Palatinate, the U.S. DOE and NSF.
\end{ack}


\end{document}